\documentclass[prd,amsmath,notitlepage,twocolumn]{revtex4-2}

\usepackage{graphicx}% Include figure files
\usepackage{dcolumn}% Align table columns on decimal point
\usepackage{bm}% bold math
\usepackage{xcolor}
\usepackage{mathrsfs}
\usepackage{url,hyperref}
\hypersetup{colorlinks,linkcolor={blue!55!black},citecolor={red!50!black},urlcolor={blue!45!black}}

\newcommand{\etal}{\textit{et al.}}

\begin{document}

\title{Alternative characterization of a black hole boundary}

\author{Pravin Kumar Dahal}
 \email{pravin-kumar.dahal@hdr.mq.edu.au}

\affiliation{
 Department of Physics and Astronomy, Macquarie University.
}
\date{\today}

\begin{abstract}
    
    After reviewing the shortcomings of existing definitions used to characterize the boundary of a black hole, we present a new method for its characterization. This definition could potentially be applied to locate the boundary of general black hole solutions of the Einstein field equations. Using our definition in general spherically symmetric spacetimes, we argue that an observer falling into a black hole encounters an infinite energy density, pressure, and flux at the black hole boundary. This result is interpreted as a firewall that prevents the growth of classical black holes. However, the possibility of black hole growth via quantum mechanical tunneling cannot be ruled out.
    
\end{abstract}


\maketitle

\section{Introduction}

Currently, the notion of trapped surfaces is widely thought of as a replacement of the event horizon to characterize a black hole boundary. This is because the event horizon is physically unobservable~\cite{1}, as one requires global knowledge of the spacetime to locate it. Moreover, event horizons are teleological in nature~\cite{2,3} and clairvoyant~\cite{4}. On the other hand, trapped surfaces are quasi-local entities requiring only the knowledge of a finite region of spacetime to locate them, which is physically possible.

The identification of trapped surfaces as black hole boundaries is straightforward and poses no problem in stationary spacetimes (where the event horizon can also be unambiguously located) and in dynamic spherically symmetric spacetimes (where the location of the event horizon is problematic). However, trapped surfaces are an ill-defined concept when we try to explore them beyond these two classes of spacetimes. This is because of two fundamental reasons: first, even in the simplest class of spacetimes (e.g.\ Schwarzschild and Vaidya), trapped surfaces depend on the choice of foliation~\cite{5,6,7} on which the family of null congruence is orthogonal. However, in spherically symmetric spacetimes, a trend is to choose the foliation that obeys the symmetry of the spacetime~\cite{8}, and this choice identifies a unique trapped surface. In axially symmetric spacetimes, such a preferred choice of foliation does not exist. Second, an arbitrary outgoing null vector is not geodesic everywhere in a generic axisymmetric spacetime. This can be seen as follows: a future-directed outgoing null vector can be written as $l_\mu=(-\alpha, \; 1, \; \beta, \; \gamma)$, where, $\alpha \ne 0$, and has three independent parameters. There are, however, four independent equations to constrain these three variables: three of the four geodesic equations $l^\mu l_{\nu; \mu}=\lambda l_\nu$ for some parameter $\lambda$ (one of the equations $l^\mu l_{r; \mu}=\lambda l_r$ is satisfied identically), and one equation from the null condition $l_\mu l^\mu=0$. These four equations will be satisfied by the three parameters of $l_\mu$ only in some regions of spacetime. However, the region where $l_\mu$ is geodesic is in general not the region where the outgoing null expansion $\theta_l$ vanishes.

More recently, the notion of a geometric horizon has been introduced as an alternative method to characterize the black hole boundary~\cite{13g,17g,15g,16g}. A conjecture is that nonstationary black holes can be defined using a hypersurface that is more algebraically special, which ensures the existence of a geometric horizon~\cite{14g,15g}. The property of being more algebraically special is foliation/observer-independent. It can be quantified invariantly by making particular combinations of curvature invariants zero at the horizon. Three different ways are pointed out in the literature to calculate the geometric horizon in general: the first approach relies on finding the suitable Killing vector field that becomes the null generator on the horizon. Then, the hypersurface where the square norm of the Killing vector vanishes is the event horizon. If it is not possible to find such a Killing vector field, then scalar polynomial curvature invariants can be calculated for stationary spacetimes and spacetimes conformal to stationary spacetimes~\cite{21g}. At the event horizon of such spacetimes, it has been shown that the squared norm of the wedge products of the $n$-linearly independent gradients of scalar polynomial curvature invariants vanishes, where $n$ is the local cohomogenity of spacetime~\cite{17g,18g}. This method of locating the black hole boundary is restricted to stationary spacetimes and their conformal transforms.

Another procedure for calculating the geometric horizon is finding the zeroes of certain combinations of Cartan invariants~\cite{19g,20g}. This method is exactly similar, in principle, to the scalar polynomial invariants for finding the geometric horizon. Thus, this method is also limited to static spacetimes and any spacetimes conformal to them. However, Cartan invariants are regarded as an improvement over scalar polynomial invariants as this method involves linear combinations of the components of the curvature tensor. Moreover, scalar polynomial invariants can be constructed from Cartan invariants.

The third procedure for the calculation of the geometric horizon is similar to the calculation of trapped surfaces, and thus, the black hole boundary calculated using this procedure may be referred to as a geometric trapped surface. The calculation is based on the conjecture that the hypersurface constituting the geometric trapped surface is more algebraically special~\cite{17g,18g}. The procedure is to identify geometrically preferred outgoing null vectors and then find the hypersurface where the expansion of such vectors vanishes. This null vector is called geometrically preferred because the congruence of the null vectors thus chosen is such that the covariant derivatives of the curvature tensor are more algebraically special there~\cite{14g,16g}. Unlike the two procedures described/mentioned above, this method of finding the geometric horizon can be extended to dynamic spacetimes as well. In particular, the geometric trapped surface was calculated for the Kerr-Vaidya spacetime in Ref.~\cite{22}, and this calculation was facilitated by the fact that the Kerr-Vaidya spacetime is Petrov II~\cite{23} and thus admits a principal null geodesic (which is algebraically special). However, the applicability of this method to more general dynamic spacetimes has not yet been tested. The equivalence/correspondence between the three procedures of calculating the geometric horizon lies in the fact that all of them involve the calculation of invariants derived from the curvature tensor (see, for example, Refs.~\cite{14g,20g} for details).

In this article, we present an alternative way of characterizing the black hole boundary and explain some of its features. In Sec.~\ref{1} we define a black hole boundary and in Sec.~\ref{2}, we apply that definition to extract some of the general features of the black hole boundary. Using our definition, we demonstrate/show that there exists a two-way barrier at the black hole boundary from which nothing can either enter or escape classically in Sec.~\ref{s3}. Finally, we discuss the consequences of our results in Sec.~\ref{4}. Here, we consider a Lorentzian manifold with the metric $g_{\mu\nu}$ of signature $\left(-,+,+,+\right)$. We use the system of units with $G=c=1$. The parameter of geodesic trajectories is denoted by $\lambda$ (which is proportional to the proper time $\tau$ for massive particles) and $\dot x= d x/d\lambda$. We also use the Einstein summation convention.

\section{Defining a new black hole boundary}\label{1}

It might be reasonable to define the black hole boundary as the hypersurface where every timelike radial geodesic approaches a null geodesic. This is the asymptotic three-surface and the underlying motivation for this classification is the classical picture of the black hole as an asymptotic state of gravitational collapse. In general, our definition demands solving the full geodesic equation to identify the black hole boundary, which might not be straightforward/feasible for general spacetimes. However, for most spacetimes, the black hole boundary according to this definition is given by the solution of $g^{rr}=0$. Exceptions are Vaidya-like spacetimes~\cite{9} for which $g^{rr}=0$ identically and $g^{t r}$ is nonzero. Coordinate transformations are possible for such spacetimes to make $g^{tr}=0$, and then the solution of $g^{rr}=0$ gives the black hole boundary.

The surface characterizing a black hole boundary is defined such that it is independent of observer/foliation generating the surface. The reason is that we are talking about the surface where every timelike radial geodesic (and not just some particular ones) becomes null. There is no general prescription for identifying the surface where every timelike geodesic approaches null. However, we will discuss below the cases where it is possible. Could this surface defined as a black hole boundary be expressed invariantly, and in particular, could the combination of curvature invariants be constructed such that they vanish at this surface? At this moment, nothing can be said for general spacetimes, and it is also unknown whether invariant characterization always implies vanishing of some combination of curvature invariants. However, we will show below that invariant characterization of such a surface is possible for spherically symmetric and Kerr spacetimes. Moreover, as this surface is foliation independent by definition, it is sufficient to show that it is coordinate independent to prove that we could characterize it invariantly.

The vector $n^\mu$ that is normal to the surface $r=\mathrm{constant}$ satisfies~\cite{35}
\begin{equation}
    g^{\mu\nu} n_\mu n_\nu= g^{r r}.
\end{equation}
Hence, $g^{r r}=0$ implies that the normal vector on the surface $r=\mathrm{constant}$ is null and that the corresponding surface is a null hypersurface. Such a hypersurface is also an event horizon~\cite{36} of Schwarzschild and Kerr spacetimes. In spherical symmetry, the normal to the surface $R= \mathrm{constant}$ can be written as $n_\mu= \nabla_\mu R$, where $R$ is the areal radius, and this allows us to express the black hole boundary invariantly as~\cite{38,39}
\begin{equation}
    g^{R R}= \nabla_\mu R \nabla^\mu R= 0.
\end{equation}

For spherically symmetric spacetimes with trajectory $l^\mu =\beta \left(1,\; \alpha, \; \gamma, \; \delta \right)$, we can take $\gamma= 0$ without loss of generality. $\delta$ can be determined from the conservation of angular momentum, and $\beta$ is fixed by the normalization condition. Neither of these quantities depend on the geodesic being either null or timelike. In such circumstances, the black hole boundary can be characterized invariantly by the surface
\begin{equation}
    l^\mu l_\mu= 0 ,
\end{equation}
for all timelike and null geodesic trajectories. In axial symmetry, $\gamma\ne 0$, in general. Although $\beta$ and $\delta$ can be taken to be the same for null and timelike trajectories, this is generally not possible for $\gamma$. Specifically, the Kerr spacetime admits three constants of motion, and two of them can be adjusted to make $\gamma$ identically the same for null and timelike geodesics. This is what makes the invariant characterization of the black hole boundary of the Kerr metric possible. This invariant surface is determined by the equation $l^\mu l_\mu= 0$ for all timelike and null geodesics. However, a complete solution of the geodesic equations of a particular spacetime under consideration might be necessary, in general.

Looking at the geodesic equations for Schwarzschild and Kerr~\cite{10} spacetimes given in Ref.~\cite{11}, we can see that all of the timelike geodesics approach null geodesics on $r-2 M =0$ and $r^2+a^2- 2r M =0$ surfaces, respectively. These surfaces are, in fact, both the event and the apparent horizon of these stationary spacetimes. In spherically symmetric spacetimes, the black hole boundary is uniquely determined~\cite{22}. We now take an example of a generic axisymmetric spacetime with five arbitrary parameters
\begin{equation}
    ds^2=g_{t t} dt^2 +2 g_{t \phi} dt d\phi +g_{r r} dr^2 +g_{\theta \theta} d\theta^2+g_{\phi \phi} d\phi^2,
    \label{n69}
\end{equation}
where the parameters $g_{\mu\nu}$ depend on the variables $t$, $r$ and $\theta$. To calculate the boundary of the black hole for this spacetime, we assume a trajectory of the form
\begin{equation}
    l^\mu =\beta \left(1,\; \alpha, \; \gamma, \; \delta \right), \label{t5}
\end{equation}
where $\beta$, $\alpha$, $\gamma$ and $\delta$ are functions of $t$, $r$ and $\theta$. Now, assuming $l^\mu$ to be a timelike trajectory, it should satisfy $l^\mu l_\mu =-1$, and this gives
\begin{equation}
    \alpha_t=\sqrt{-\frac{1}{g_{r r}}\left(\frac{1} {\beta^2}+g_{t t}+ 2 \delta g_{t \phi} +\gamma^2 g_{\theta \theta}+ \delta^2 g_{\phi \phi}\right)}.
    \label{r60}
\end{equation}
This timelike radial trajectory coincides identically with the null trajectory given by
\begin{equation}
    \alpha_n= \sqrt{-\frac{1}{g_{r r}}\left(g_{t t}+ 2 \delta g_{t \phi} +\gamma^2 g_{\theta \theta}+ \delta^2 g_{\phi \phi}\right)},
    \label{ngdesic4}
\end{equation}
on the surface $g^{r r}=g_{r r}^{-1}=0$, irrespective of the form of the parameters $g_{\mu \nu}$, $\gamma$, and $\delta$, assuming that the normalization factor $\beta \ne 0$ there (this is to be expected as otherwise the trajectory is identically zero at $g^{r r}=0$, appendix~\ref{aa}). Thus, for the general form of an axisymmetric metric given by Eq.~\eqref{n69}, adopting the definition presented here gives $g^{rr}=0$ as the black hole boundary. The general definition of the black hole boundary should comply with our common intuition that nothing can escape from inside of the black hole. Our definition indeed agrees with this classical intuition, as will be explained below.

The spacetime should not necessarily be the solution of the Einstein field equation to apply our definition to determine the black hole boundary. Let an object be moving in a straight line with constant acceleration in Minkowski spacetime. It is interesting to note that the virtual surface where its trajectory becomes null is a black hole boundary by our definition. We will study this new consequence in more detail in the near future (see also Rindler horizon~\cite{14,37}). Regarding the existence of this surface in spacetimes that do not have a trapped region (or event horizon), nothing can be said in general. However, for the particular case of spacetimes that are conformal transforms of the Schwarzschild black hole, if the event horizon exists, then the black hole boundary characterized by our definition exists. This is because the event horizon, being null, is invariant under conformal transformation, and this surface satisfies the criterion to be the black hole boundary by our definition. We have to examine particular cases for more general spacetimes to see if this is possible.

\section{Features of the new black hole boundary}\label{2}

Let us simplify Eq.~\eqref{ngdesic4} further by using the equation of motion in the $\phi$-direction
\begin{equation}
    \frac{d}{d\tau}\bigg(\frac{\partial\cal{L}}{\partial \dot{\phi}}\bigg)-\frac{\partial\cal{L}}{\partial \phi}=0,
\end{equation}
where ${\cal{L}}=- d\tau^2/2$ is the Lagrangian density. As the Lagrangian is independent of $\phi$, we get
\begin{equation}
    \frac{\partial\cal{L}}{\partial \dot{\phi}}= g_{t\phi}\dot t+ g_{\phi\phi}\dot \phi=\text{constant}=L,
    \label{pe6}
\end{equation}
where $L$ is an arbitrary constant. We thus have
\begin{equation}
    \delta=\frac{L-g_{t\phi}\beta}{g_{\phi\phi}\beta}.
\end{equation}
Next, we substitute this expression into Eq.~\eqref{ngdesic4} to obtain
\begin{equation}
    \alpha=\sqrt{-\frac{g^{r r}}{\beta^2 g_{\phi\phi}}\left(g_{t t}g_{\phi\phi}\beta^2-g_{t\phi}^2\beta^2+\gamma^2\beta^2 g_{\theta\theta} g_{\phi\phi}+L^2\right)},
    \label{re8}
\end{equation}
where $\alpha=\alpha_t=\alpha_n$ near the black hole boundary $g^{r r}=0$. In the limit $g^{r r}\to 0$ we have $g^{r r}L^2\to 0$, and thus Eq.~\eqref{re8} reduces further to
\begin{equation}
    \alpha=\sqrt{-g^{r r}\left(\frac{g_{t t}g_{\phi\phi}-g_{t\phi}^2}{g_{\phi\phi}}+\gamma^2 g_{\theta\theta} \right)}.
\end{equation}
At this point, we can make two fundamental assumptions that hold in Schwarzschild and Kerr black hole solutions of general relativity. First, we assume that the radial and azimuthal trajectories $\dot r=\beta\alpha$ and $\dot \theta=\beta\gamma$, respectively, are finite. Second, we assume that the metric coefficient $g_{\theta\theta}$ is finite at the black hole boundary.
These assumptions are satisfied in known cases, and are expected to hold in more general black hole solutions of the Einstein equations as well. The implication of these assumptions near $g^{r r}\to 0$ is $\gamma^2\beta^2 g^{r r}g_{\theta\theta}\to 0$, which gives
\begin{equation}
    \dot r=\beta\alpha=\sqrt{-g^{r r} \frac{g_{t t}g_{\phi\phi}-g_{t\phi}^2}{g_{\phi\phi}}\beta^2},
    \label{rv10}
\end{equation}
and
\begin{equation}
    \frac{dr}{dt}=\alpha=\sqrt{-g^{r r} \frac{g_{t t}g_{\phi\phi}-g_{t\phi}^2}{g_{\phi\phi}}}.
    \label{rv11}
\end{equation}
We will apply these relations to a particular example below.

\subsection{An example of the stationary Kerr metric}

To calculate the radial velocity of an infalling particle (either massive or massless) given by Eq.~\eqref{rv10} and Eq.~\eqref{rv11}, let us consider the case of the stationary Kerr metric
\begin{multline}
    ds^2=-\bigg(1-\frac{2 M r}{\rho^2}\bigg)dt^2-\frac{4 a M r \sin^2\theta}{\rho^2}dt d\phi
    +\frac{\rho^2}{\Delta}dr^2\\
    +\rho^2 d\theta^2+\frac{(r^2+a^2)^2-a^2 \Delta \sin^2\theta}{\rho^2}\sin^2\theta d\phi^2,
\end{multline}
where $\rho^2=r^2+a^2 \cos^2\theta$ and $\Delta=r^2+a^2-2 M r$. At first glance, the finiteness of the values of $g_{tt}$, $g_{t\phi}$, and $g_{\phi\phi}$ gives $dr/dt=0$ (from Eq.~\eqref{rv11}), which is one of the remarkable features at the black hole boundary. 
This implies that an infalling particle (either massive or massless) takes an infinite amount of time to cross the boundary of a black hole in the frame of a distant observer.

Again, to calculate $\dot r$, let us first calculate $\beta=\dot t$, which is done using the equation of motion
\begin{equation}
    \frac{d}{d\tau}\bigg(\frac{\partial\cal{L}}{\partial \dot{t}}\bigg)-\frac{\partial\cal{L}}{\partial t}=0,
\end{equation}
where ${\cal{L}}$ is the Lagrangian density. For the Kerr metric, the Lagrangian is independent of $t$, which gives
\begin{equation}
    \frac{\partial\cal{L}}{\partial \dot{t}}= g_{tt}\dot t+ g_{t\phi}\dot \phi=\text{constant}=E,
\end{equation}
where $E$ denotes an arbitrary constant. Combined with Eq.~\eqref{pe6}, this equations gives
\begin{equation}
    \beta=\frac{E g_{\phi\phi}-L g_{t\phi}} {g_{\phi\phi} g_{tt}- g_{t\phi}^2}. \label{b18}
\end{equation}
Finally, substituting the values of the metric coefficients into Eq.~\eqref{rv10} gives
\begin{equation}
    \dot r=\frac{(r^2+a^2) E+ a L}{\rho^2}, \label{re16}
\end{equation}
which exactly matches the general geodesic equation in Kerr spacetime (for both massive and massless particles) given in Ref.~\cite{11} at the horizon $\Delta=0$. Similarly, from Eq.~\eqref{rv11}, we get
\begin{equation}
    \frac{dr}{dt}=\pm\frac{\Delta}{\sqrt{(r^2+a^2)^2-a^2 \Delta \sin^2\theta}}.
\end{equation}
For a general spherically symmetric spacetimes and for stationary axisymmetric spacetimes at least, our definition of the black hole boundary thus implies another feature, which is vanishing of the radial velocity of infalling objects (both massive and massless) in the frame of a distant observer. This feature was first noticed in the explicit calculation of the radial geodesic equations in the Schwarzschild and Kerr spacetimes~\cite{11}. To summarize, a black hole boundary is a surface where (i) the radial velocity of massive particles approaches that of light in vacuum in the frame of a comoving observer (at least in general spherically symmetric spacetimes where purely radial motion is possible); (ii) the radial velocity of both massive and massless particles approaches zero in the frame of a distant noninertial observer (at least in all spherically symmetric and in stationary axisymmetric spacetimes).
These features are a direct consequence of the definition that all timelike radial geodesics approach null geodesics at the black hole boundary. From here on, we will restrict ourselves to the case of spherical symmetry. This restriction allows us to assume a purely radial geodesic trajectory, which becomes null upon reaching the black hole boundary.

\section{Barrier at the black hole boundary}\label{s3}

We first consider the case of a massive particle as it travels toward a black hole boundary. Our definition above implies that the black hole boundary is a surface where the trajectory of massive particles approaches that of the null trajectory (at least in spherical symmetry, where a purely radial trajectory is possible). Although general relativity does not restrict massive particles from moving at the velocity of light, special relativity (cf. Lorentz transformation) simply does not allow this. This phenomenon, from the Einstein mass relation $m=m_0/\sqrt{1-v^2}$, results in an infinite density and pressure at the black hole boundary according to a comoving observer. This divergence of local density and pressure is called a firewall and is a consequence of careful analysis of the classical black hole boundary rather than from the semiclassical approximation~\cite{12} or the information loss paradox~\cite{18} (see Sec.~\ref{4b} for a more detailed explanation of the firewall).

We have concluded previously that massive particles fall inside of the black hole boundary in a finite proper time. However, in reaching this conclusion, we have completely ignored the fact that, in the frame of a comoving observer, massive particles become massless at the boundary. One of the implications of this is the firewall, as mentioned in the above paragraph. We discuss another consequence here, where we consider the case of a massless particle as it travels toward a black hole boundary. The concept of proper time is not useful to describe massless particles. To illustrate this point, let us imagine an observer comoving with massless particles that radially fall toward the black hole boundary. The proper time elapsed is zero for the comoving observer. However, the geodesic equation for massless particles can be written as $d\lambda= dr/E$ using Eq.~\eqref{re16}, where $a=0$ in spherical symmetry and $L=0$ for radial geodesics. The parameter $\lambda$ must not be interpreted as the proper time $\tau$ for massless particles as $\Delta\tau=0$, and $\Delta\lambda$ is not always zero. Since every particle becomes massless as it reaches the black hole boundary, the statement that particles cross the horizon in a finite proper time could be wrong. It is fair to say that we were completely unaware of the hurdle a particle has to face at the boundary in reaching the interior of a black hole.

Even though the notion of proper time is not useful/sensible to describe the crossing of a black hole boundary, the coordinate time of a distant observer works well to describe this phenomenon. The result that it takes an infinite amount of time for massless particles to cross a black hole horizon in the frame of a distant observer is due to the increase in the spatial distance caused by spacetime curvature. In Schwarzschild spacetime, the actual radial distance massless particles have to travel due to the increase caused by spacetime curvature can be calculated as the change in $r^*$, where
\begin{equation}
    r^*=r+2 M \ln \left(\frac{r}{2 M}-1\right),
\end{equation}
is the tortoise coordinate. Indeed, $r^*$ gives the actual distance, as seen from the fact that the velocity of light is one when the distance is calculated with respect to $r^*$, that is,
\begin{equation}
    \frac{d r^*}{d t}=\frac{d r^*}{d r}\frac{d r}{d t}=1.
\end{equation}
Here, we have used the radial equation of motion for a massless particle in Schwarzschild spacetime (see, for example, Ref.~\cite{11}).

%We can establish a perfect analogy of light traveling in the Schwarzschild spacetime by considering light traveling in a spherically symmetric medium of ever-increasing density/refractive index towards the center. A region of a high gravitational field is the region of a high refractive index. So, when light propagates into high density, it gets bend towards to center, simply following the principle of least time. This bending happens at each point because of continuously varying gravitational field/refractive index until the light reaches the surface with infinite density (provided that the impact parameter is small enough). This surface is the black hole boundary. There is no such medium with an infinite refractive index, but we expect such a high gravitational field to occur in the Universe. Upon traveling into this medium, a distant observer sees the velocity of light varying at each point, ultimately going to zero $d r/d t\to 0$ at the surface of an infinitely high refractive index. However, in reality, the light is travelling with a constant velocity $c=d r^*/d t=1$ towards the region of ever increasing optical path length:
%\begin{equation}
    %d r^*=\frac{1}{1-\frac{2 M}{r}}d r
%\end{equation}
If the gravitational adage ``what goes in, must come out" is true, then its contrapositive ``if nothing can come out, nothing can go in" is also true.

\subsection{More comments on the expansion of null geodesics}

The expansion of a null geodesic is a measure of divergence/convergence of the vector field that is tangent to the null geodesic. In the region where the null geodesic neither diverges nor converges, its expansion is zero. To see this, let us consider a future-directed field of null vectors $l^\alpha$ in the orientable spacetime manifold ${\cal M}$. There exists an auxiliary null field $n^\alpha$, which is not unique, but satisfies $l^\alpha n_\alpha=-1$. The divergence (or the convergence) of the field of null congruences $l^\mu$ is given by its covariant derivative $l^\mu_{;\mu}$. If the vector field $l^\mu$ is arbitrarily parametrized, then we affinely parametrize it before the calculation of its divergence, which introduces the additional correction term $\kappa=-n_\mu l^\nu l^\mu_{;\nu}$, called surface gravity. Thus, the divergence (or convergence) of the affinely parametrized null congruence is given by~\cite{14}
\begin{equation}
    \theta_l=l^\mu_{;\mu}+n_\mu l^\nu l^\mu_{;\nu},
    \label{e1}
\end{equation}
where $\theta_l$ denotes the expansion (more details on the expansion of congruences are provided in the textbook by Poisson~\cite{15}).

As an example, we take the Schwarzschild spacetime (which can easily be generalized to all spherically symmetric spacetimes by substituting $M\to M(t,r)$ and $g_{t t} \to e^{-2 h(t,r)} g_{t t}$), where the null vector field providing the preferred choice of foliation is radial ~\cite{8}. We consider the radial null geodesic $l^\mu$ which is outgoing in the sense that it points away from the trapped region,
\begin{align}
    l_{\mu}=\left(-\left(1-\frac{2 M}{r}\right),1, 0, 0 \right),\\
    n_{\mu}=\frac{1}{2}\left(-1,-\frac{1}{\left(1-\frac{2 M}{r}\right)}, 0, 0 \right).
\end{align}
The expansion for the field $l^\mu$ is $\theta_l= \frac{2}{r}\left(1-\frac{2 M}{r}\right)$, which vanishes at $r=2 M$. Similarly, if we take the radial null geodesic $l^\alpha$, which is ingoing in the sense that it points toward the trapped region,
\begin{align}
    l_{\mu}=\left(-\left(1-\frac{2 M}{r}\right),-1, 0, 0 \right),\\
    n_{\mu}=\frac{1}{2}\left(-1,-\frac{1}{\left(1-\frac{2 M}{r}\right)}, 0, 0 \right),
\end{align}
then the expansion for this field is $\theta_l= -\frac{2}{r}\left(1-\frac{2 M}{r}\right)$, which again vanishes at $r=2 M$. Notice the positive sign for the divergence and the negative sign for the convergence of the vector field.

Although expansion scalars are coordinate independent, they do depend on the position of the observers (see Refs.~\cite{5,6,7,8} for a discussion of the observer/foliation dependence of trapped surfaces). This is why calculation of the expansion scalars of null geodesics should not be taken seriously in regard to locating/identifying/determining the black hole boundary~\cite{22}. One of the results involving expansion scalars that could be taken seriously is the focusing theorem, which implies that gravitation always converges provided the strong energy condition holds~\cite{15,17}. The reason for the divergence/convergence of both the incoming/outgoing null geodesics being zero at the black hole boundary has to do with the infinite spatial extension of spacetime outside of the black hole in the frame of an infalling observer (or infinite temporal extension in the frame of a distant observer).

\subsection{Comments on the firewall}\label{4b}

The possibility that an infalling observer encounters a firewall at the black hole boundary was first put forward in 2012 by Almheiri \etal~\cite{18} to resolve an apparent black hole complementarity problem~\cite{19,20}. In a completely different context, it was shown that an infalling observer toward expanding spherically symmetric black holes perceives a firewall at the boundary~\cite{12}. This firewall is an outcome of the assumptions of finite time formation and regularity of the trapped surface. This work was later extended to axially symmetric black hole solutions by considering a particular Kerr-Vaidya metric, thereby showing that the firewall is not an artifact of spherical symmetry~\cite{21,16}. The firewall discussed above is similar in nature to both of these firewalls,
%but is more fundamental \tcb{[in what sense?]} and \tcb{\st{is}} 
but of classical origin rather than of semiclassical or quantum origin.

Let us assume a massive observer with energy density $m_0$ infalling toward the Schwarzschild black hole (as pointed out above, this situation can easily be generalized to all spherically symmetric black hole solutions). If the pressure is assumed to be zero, then the only nonzero component of the energy-momentum tensor of an observer in its rest frame is
\begin{equation}
    T^0_0=-\rho_0 \quad \textrm{or} \quad T^{0 0}= -\frac{\rho_0}{g_{t t}}.
\end{equation}
Next, we calculate the energy density of an observer in a frame moving with velocity $u^\mu=(\dot{T},\; \dot{R},\; 0,\; 0)$ using the relation $\rho=T_{\mu\nu}u^\mu u^\nu$, where $u^\mu$ is the four-velocity that satisfies $u^\mu u_\mu= -\chi$ for $0 \leq \chi\leq 1$. This gives
\begin{equation}
    \dot{T}^2= -\frac{\chi+ g_{r r} \dot R^2}{g_{t t}}.
\end{equation}
We thus have
\begin{equation}
    \rho= T_{0 0}{u^0}^2= \rho_0 \left(\chi+ g_{r r} \dot{R}^2 \right).
\end{equation}
From the definition given in Sec.~\ref{1}, $\chi \to 0$ upon approaching the black hole boundary, and from Eq.~\eqref{re16} we see that $\dot R$ is finite everywhere, thereby giving
\begin{equation}
    \rho \to g_{r r} \rho_0 \dot R^2
\end{equation}
on approach to the black hole boundary. Thus, the energy density of an infalling observer diverges at the black hole boundary. We can also calculate the pressure $p=T_{\mu\nu}n^\mu n^\nu$ perceived by an infalling observer, where $n^\mu$ is an outgoing spacelike normal satisfying $n^\mu u_\mu=0$ and $n^\mu n_\mu>0$. Such normals are not uniquely determined, and one of them can be written as
\begin{equation}
    n_\mu= \sqrt{- g_{t t} g_{r r}} (-\dot R, \dot T,0,0).
\end{equation}
We thus have
\begin{equation}
    p= T_{0 0} {n^0}^2= \rho_0 g_{r r} \dot R^2,
\end{equation}
thereby showing that the pressure also diverges at the black hole boundary. The flux $\varphi=T_{\mu\nu}u^\mu n^\nu$ for the infalling observer is given by
\begin{equation}
    \varphi= T_{0 0} n^0 u^0 \to \pm \rho_0 g_{r r} \dot R^2,
\end{equation}
where the positive sign corresponds to an outgoing observer and the negative sign to an ingoing observer. Therefore, at the black hole boundary, the energy density, pressure and flux of an infalling observer all diverge:
\begin{equation}
    \rho= p= \pm \varphi= \rho_0 g_{r r} \dot R^2.
\end{equation}
This firewall is indeed not an artifact of a coordinate system (see appendix~\ref{appa}). However, it is observer dependent and can be seen only by observers infalling at the black hole boundary. Although an infalling observer will register an infinite increase in energy density, pressure and flux, tidal force in his/her reference frame remains finite (as the curvature invariant is finite) when approaching the boundary. The singularity of this type is known for a very long time and is called weak nonscalar, whimper singularity~\cite{40}. They appeared in the classical models of the inner horizon of the charged Vaidya spacetime~\cite{36}.

\section{Discussion and conclusion} \label{4}

We have presented a new method for identifying the boundary of a black hole. Although in general solving the full radial geodesic equation of the spacetime is necessary to characterize the black hole boundary, for some spacetimes, it simply corresponds to the solution of $g^{rr}=0$. Eq.~\eqref{n69} is an example of such a of spacetime. The validity of our definition for characterizing the black hole boundary can be deduced from the fact that such a surface exists in the black hole solutions of the Einstein equations (for example, the Schwarzschild and the Kerr solution). The universality of our definition relies on whether a surface where all timelike radial geodesics approaching null geodesics exists in all black hole solutions.

Although it is argued that nothing can enter the black hole classically, this possibility is not excluded quantum mechanically due to the existence of an infinite barrier at the black hole boundary where a law of special relativity is violated in the frame of a comoving observer. If we assume that it is indeed impossible to cross this barrier classically (that is, if we assume that, for a comoving observer, the violation of special relativity does not occur at the horizon), then growth of the black hole must be accompanied by quantum mechanical tunneling. The finite width barrier that is required for tunneling could be constructed in a subtle way: as the particle falls into the black hole, its mass and thus its radius increase. It is this increase in the size of the black hole that sets the scale for the tunneling problem. The Parikh-Wilczeck tunneling approach~\cite{33,34} can be used to calculate the tunneling probability $\Gamma=\exp \left(-2 \text{Im} I\right)$, where $\text{Im} I$ is the imaginary part of the action for the classically forbidden trajectory. The problem, however, is that we do not know the interior black hole spacetime inside of the barrier at which tunneling occurs. The possibility of black hole growth by quantum tunneling will be explored in future works.

\appendix

\section{Geodesic equation in time coordinate} \label{aa}

In general spacetimes, we do not expect $\beta$ to be identically zero at the black hole boundary. For the trajectory $l^\mu$ of Eq.~\eqref{t5}, $l^\mu l_\mu=0$ at the black hole boundary implies
\begin{equation}
    \beta^2 \left( g_{t t}+ 2 \delta g_{t \phi} + \alpha^2 g_{r r}+ \gamma^2 g_{\theta \theta}+ \delta^2 g_{\phi \phi}\right)= 0.
\end{equation}
For this to hold, either $\beta=0$, or
\begin{equation}
    \left( g_{t t}+ 2 \delta g_{t \phi} + \alpha^2 g_{r r}+ \gamma^2 g_{\theta \theta}+ \delta^2 g_{\phi \phi}\right)= 0, \nonumber
\end{equation}
or both. In particular, $\beta=0$ implies $l^\mu = \left(0,\; 0, \; 0, \; 0 \right)$ at the black hole boundary. This is very unlikely to happen in general spacetimes as it is not even true for stationary spacetimes, as will be shown below.

The equation of motion in the $t$-component is
\begin{equation}
    \frac{d}{d\tau}\left(\frac{\partial\cal{L}}{\partial \dot{t}}\right)-\frac{\partial\cal{L}}{\partial t}=0,
\end{equation}
${\cal{L}}$ being the Lagrangian density. For the metric of Eq.~\eqref{n69}, this reduces to
\begin{equation}
    \frac{d}{d\tau}\big(2 g_{tt} \dot t+ 2 g_{t\phi} \dot\phi\big)-\frac{1}{2}\frac{\partial g_{\mu\nu}}{\partial t} \dot x^{\mu} \dot x^{\nu}=0. \label{a22}
\end{equation}
From Eq.~\eqref{t5}, we can see that $\dot t= \beta$, the normalization factor. It is very hard to solve Eq.~\eqref{a22}, in general. However, the exact solution could be obtained for the general stationary metric, and the explicit form of $\beta$ in that case is given in Eq.~\eqref{b18}. Surely, for stationary spacetimes, constants of motion $E$ and $L$ (which depend on the initial conditions of the trajectory) can be adjusted such that $\beta$ is nonzero at the black hole boundary.

\section{Firewall in advanced coordinates} \label{appa}

To show that the firewall we have obtained in Sec.~\ref{4b} is not an artifact of a particular coordinate system, we consider the Schwarzschild metric in advanced coordinates~\cite{13}
\begin{equation}
    ds^2= -\left(1-\frac{2 M}{r}\right) dv^2+ 2 dv dr+ r^2 d\theta^2+ r^2 \sin^2\theta d\phi^2
\end{equation}
as an example. Our prescription for finding the black hole boundary of this spacetime involves solving the radial geodesic
\begin{equation}
    \dot r^2= \left(1-\frac{2 M}{r}\right) \left(\chi-\frac{L}{r^2}\right)+ E^2, \label{a2}
\end{equation}
where $L$ and $E$ are constants of motion and $\chi= 0$ for null geodesics and $0<\chi \leq 1$ for timelike geodesics. The timelike radial trajectory becomes null at $r= 2M$, which is the black hole boundary of this spacetime. Now, as in Sec.~\ref{4b}, we can calculate the energy density, pressure and flux.

We calculate the energy density, pressure and flux of an observer moving in a frame with velocity $u^\mu=(\dot{V},\; \dot{R},\; 0,\; 0)$. The four-velocity $u^\mu$ satisfies $u^\mu u_\mu= -\chi$ for $0 \leq \chi\leq 1$ and this gives
\begin{equation}
    \dot V= \frac{\dot R+ \sqrt{\dot R^2+\left(1-\frac{2 M}{r}\right)\chi}}{1-\frac{2 M}{r}}.
\end{equation}

We thus have
\begin{equation}
    \rho= T_{0 0}{u^0}^2= \rho_0 \frac{\left(\dot R+ \sqrt{\dot R^2+\left(1-\frac{2 M}{r}\right)\chi}\right)^2}{1-\frac{2 M}{r}}.
\end{equation}
As $\chi \to 0$ on approaching the black hole boundary, and from Eq.~\eqref{a2} $\dot R$ is finite everywhere, thereby giving
\begin{equation}
    \rho \to \frac{4 \rho_0 \dot R^2}{1-\frac{2 M}{r}},
\end{equation}
on approach to the black hole boundary. Thus, the energy density of an infalling observer diverges at the black hole boundary. Similarly, we can show that the pressure and flux of the infalling observer also diverges at the black hole boundary. It is straightforward to generalize this result to general spherically symmetric metrics in advanced and retarded coordinates. This clearly illustrates that the firewall is not an artifact of a particular coordinate system. However, the illustration is not independent of a coordinate system (we have chosen coordinates with a nonsingular metric).

\begin{acknowledgments}
I want to thank Sebastian Murk and the anonymous referee for their comments and suggestions in improving this article. Pravin Kumar Dahal is supported by an International Macquarie Research Excellence Scholarship.
\end{acknowledgments}

\end{document}